\documentclass[english,reprint,aps,prl,superscriptaddress]{revtex4-1}
\usepackage[T1]{fontenc}
\usepackage[latin9]{inputenc}
\usepackage{color}
\usepackage{babel}
\usepackage{amsmath}
\usepackage{amssymb}
\usepackage{ulem}
\usepackage[colorlinks=true,linktocpage=true,linkcolor=blue,citecolor=blue]{hyperref}

\makeatletter
\usepackage{babel}

\makeatother

\begin{document}
\title{Ideal spin hydrodynamics from Wigner function approach}
\author{Hao-Hao Peng}
\affiliation{Interdisciplinary Center for Theoretical Study and Department of Modern
Physics, University of Science and Technology of China, Hefei, Anhui
230026, China}
\affiliation{Peng Huanwu Center for Fundamental Theory, Hefei, Anhui 230026, China}
\author{Jun-Jie Zhang}
\affiliation{Northwest Institute of Nuclear Technology, Xi'an 710024, China}
\author{Xin-Li Sheng}
\affiliation{Key Laboratory of Quark and Lepton Physics (MOE) and Institute of
Particle Physics, Central China Normal University, Wuhan, 430079,
China}
\author{Qun Wang}
\affiliation{Interdisciplinary Center for Theoretical Study and Department of Modern
Physics, University of Science and Technology of China, Hefei, Anhui
230026, China}
\affiliation{Peng Huanwu Center for Fundamental Theory, Hefei, Anhui 230026, China}

\begin{abstract}
Based on the Wigner function in local equilibrium, we derive hydrodynamical quantities
for a system of polarized spin-1/2 particles: the particle number current density,
the energy-momentum tensor, the spin tensor, and the dipole moment tensor. Comparing with
ideal hydrodynamics without spin, additional terms at first and second order in the Knudsen number $\text{Kn}$ and the average spin polarization $\chi_s$
have been derived. The Wigner function can be expressed in terms of matrix-valued distributions,
whose equilibrium forms are characterized by thermodynamical parameters in quantum statistics.
The equations of motions for these parameters are derived by conservation laws at the leading and
next-to-leading order $\text{Kn}$ and $\chi_s$.
\end{abstract}

\maketitle

\paragraph{Introduction. }

Large orbital angular momenta (OAM) perpendicular to the reaction plane can be generated
in noncentral relativistic heavy ion collisions.
Part of initial OAM are converted to spin angular momenta of quarks
through the spin-orbit coupling \citep{Liang:2004ph,Voloshin:2004ha},
similar to the Einstein-de Hass and Barnett effects
\citep{Einstein:1915,Barnett:1935}. At the freeze-out stage, these
polarized quarks combine to form polarized hadrons.
The polarization of hyperons along the global OAM (global spin polarization)
has recently been measured in the STAR experiment \citep{STAR:2017ckg,Adam:2018ivw}.
According to quantum statistics \citep{Becattini:2007sr,Becattini:2013vja,Becattini:2013fla},
the spin polarization is connected to the thermal vorticity
$\omega_{\mu\nu}^\text{th}\equiv-(\partial_{\mu}\beta_{\nu}-\partial_{\nu}\beta_{\mu})/2$,
where $\beta_{\mu}=u_{\mu}/T$ is the ratio of the flow velocity $u_{\mu}$
to the temperature $T$. Hydrodynamical and transport
models \citep{Karpenko:2016jyx,Xie:2017upb,Li:2017slc,Sun:2017xhx,Wei:2018zfb}
can describe the experimental data of the global $\Lambda$ polarization.
However, the longitudinal polarization of $\Lambda$s calculated in hydrodynamical
and transport models \citep{Becattini:2017gcx} has an opposite sign to experimental results
\citep{Adam:2019srw}, which is called a ``sign problem''.
Several attempts have been made in explaining the ``sign problem'',
including different choices of spin potentials \citep{Wu:2019eyi},
non-relativistic connection between the polarization and vorticity \citep{Florkowski:2019voj},
off-equilibrium contributions according to transport model simulations \citep{Liu:2019krs},
the contribution from the shear-tensor \citep{Liu:2021uhn,Fu:2021pok,Becattini:2021suc,Becattini:2021iol},
and the equation of states and freeze-out temperatures \citep{Yi:2021ryh},
but at present no definite conclusion has been drawn.
Meanwhile, it is expected to
exist a parity violation in a hot QCD matter, which may be observed through the Chiral Magnetic Effect (CME) \citep{Kharzeev:2004ey,Kharzeev:2007jp,Fukushima:2008xe}. Isobar collisions have been carried out by the STAR collaboration \citep{Voloshin:2010ut,Wang:2018ygc,STAR:2019bjg} but no CME signatures have been observed \citep{STAR:2021mii}. In order to better understand the ``sign problem'' and search for the CME signature, more theoretical efforts and numerical simulations for the dynamical evolution of spin degrees of freedom are needed, see e.g. Refs. \citep{Liu:2020ymh,Gao:2020vbh,Gao:2020lxh,Gao:2020pfu} for recent reviews of spin effects in heavy ion collisions.

One way to describe the evolution of spin degrees of freedom in the quark gluon plasma is through relativistic spin hydrodynamics.
Relativistic viscous hydrodynamics (without spin) is a successful framework to describe the evolution of strong interaction matter in heavy-ion collisions and therefore is widely used in numerical simulations \citep{Romatschke:2009im,Heinz:2013th,Gale:2013da,Romatschke:2017ejr,Shen:2020mgh,Wu:2021xgu}.
Comparing with the spinless case, the spin hydrodynamics
includes an additional conservation law for the total angular momentum,
which allows a conversion between spin and OAM. The spin hydrodynamics
was proposed many years ago \citep{Weyssenhoff:1947iua}.
Recent developments in the relativistic hydrodynamics include the effective theory approach \citep{Becattini:2009wh,Montenegro:2017rbu,Montenegro:2017lvf},
the derivation using the local equilibrium distribution functions \citep{Florkowski:2017ruc,Florkowski:2017dyn,Florkowski:2018fap,Florkowski:2019fmz}
introduced in \citep{Becattini:2013fla}.
The second order hydrodynamical equations can be derived in many approaches,
such as the generating function method \citep{Gallegos:2021bzp}, relaxation
time approximation \citep{Bhadury:2020cop,Bhadury:2020puc}, and entropy
current analysis based on the second law of thermodynamics
\citep{Hattori:2019lfp,Fukushima:2020ucl,Li:2020eon,She:2021lhe}.

In this work, we will derive relativistic spin hydrodynamics for ideal fluids
from Wigner functions. The conventional ideal hydrodynamics
can be recovered in the spinless case.
The Wigner function is expressed in terms of matrix-valued distributions,
whose local equilibrium forms are given by quantum statistics \citep{Becattini:2013fla}.
We first obtain the Wigner function up to the second order in space-time derivatives
by solving the equation of the Wigner function.
From the Wigner function, we derive hydrodynamical quantities as functions
of thermodynamical parameters: the temperature, the flow velocity,
the chemical potential and the spin potential. The equations
of motions for these parameters are obtained from conservation laws.

Throughout this work, we use natural units $\hbar=c=k_{B}=1$.
The antisymmetric tensor is defined as $A^{[\mu}B^{\nu]}\equiv(A^{\mu}B^{\nu}-A^{\nu}B^{\mu})/2$,
while the symmetric one is defined as $A^{\{\mu}B^{\nu\}}\equiv(A^{\mu}B^{\nu}+A^{\nu}B^{\mu})/2$.

\vspace{1.5em}

\paragraph{Power counting. \label{sec:Power-Counting}}
Before we start our discussion, we want to clarify the power counting scheme used throughout this letter. In an ordinary Navier-Stokes type of fluid, it is usually assumed that the hydrodynamical limit can be reached when microscopic details in the system can be safely neglected. This limit is quantified as $\text{Kn}\equiv l/L\ll 1$, where the Knudsen number $\text{Kn}$ is defined as the ratio of the typical microscopic length scale $l$ to the macroscopic length scale $L$. In our discussion, $l$ is associated with the microscopic mean-free-path, while $L$ is described by $1/L\sim \left|\partial_\mu O/O\right|$, the typical scale of the gradient of the macroscopic quantity $O$, such as the charge and energy densities, etc. Therefore a gradient expansion is equivalent to an expansion with respect to the Knudsen number.

For systems with spin degrees of freedom, we introduce an additional parameter $\chi$ to quantify spin polarization. This parameter can be given by either the average spin polarization per particle or the average magnetic dipole moment,
\begin{equation}
\chi_s\sim\frac{\left|\mathcal{S}^{\lambda,\mu\nu}\right|}{n}\sim\frac{\left|\mathcal{M}^\mu\right|}{n}\,,
\end{equation}
where $\mathcal{S}^{\lambda,\mu\nu}$ is the spin tensor density, $\mathcal{M}^\mu$ is the magnetic dipole moment density, and $n$ is the particle number density. If spin effects are induced by the thermal vorticity as in global equilibrium, we expect $\chi_s\sim\text{Kn}$ and therefore spin effects are at least first order in the Knudsen number. In this letter we consider a weakly polarized system by demanding $\chi_s\lesssim \text{Kn}$ with two expansion parameters $\chi_s$ and $\text{Kn}$. For physical observables, we keep zeroth order terms $\mathcal{O}(1)$, first order terms $\mathcal{O}(\text{Kn})$, $\mathcal{O}(\chi_s)$, as well as second order terms $\mathcal{O}(\text{Kn}^2)$, $\mathcal{O}(\text{Kn}\,\chi_s)$, and $\mathcal{O}(\chi_s^2)$.

\paragraph{Wigner function in thermal equilibrium. \label{sec:Wigner-function-at}}

The distribution function $f(t,{\bf x},{\bf p})$ for a classical particle
is the probability of finding a particle with the momentum ${\bf p}$ at
the space-time point $(t,{\bf x})$. However, the position and the momentum of
a quantum particle can not be determined simultaneously according to the uncertainty principle.
Thus the quantum analogue of $f(t,{\bf x},{\bf p})$ is
the covariant Wigner function defined as a two-point correlation function \citep{Heinz:1983nx,Elze:1986qd}
\begin{equation}
W(x,p)\equiv\int\frac{d^{4}y}{(2\pi)^{4}}e^{-ip\cdot y}\left\langle :\bar{\psi}\left(x+\frac{y}{2}\right)\otimes\psi\left(x-\frac{y}{2}\right):\right\rangle \,,\label{eq:definition_Wigner}
\end{equation}
where $\left\langle :\ :\right\rangle $ denotes the ensemble average
and $\otimes$ denotes the tensor product of two matrices, i.e., $[A\otimes B]_{\alpha\beta}=A_{\beta}B_{\alpha}$.
One can verify that $x^{\mu}$ is commutable with $p^{\mu}$ because
$x^{\mu}$ is the central position of two points, while $p^{\mu}$ is
the momentum conjugate to the relative distance between two points.
This indicates that the Wigner function is well-defined in phase space.
Note that we neglected gauge fields such as electromagnetic fields
throughout this paper. If a gauge field is considered, one should
put a gauge link between $\bar{\psi}\left(x+y/2\right)$ and $\psi\left(x-y/2\right)$
as in Ref. \citep{Heinz:1983nx,Vasak:1987um}.

The kinetic equation of the Wigner function can be
derived from the Dirac equation \citep{Elze:1986qd,Vasak:1987um},
\begin{equation}
\left[\gamma_{\mu}\left(p^{\mu}+\frac{i}{2}\partial^{\mu}\right)-m\right]W(x,p)=0\,.\label{eq:equation_Wigner}
\end{equation}
At zeroth order in gradient, the Wigner function has the following solution,
\begin{align}
W_0(x,p) & =\frac{1}{(2\pi)^{3}}\delta(p^{2}-m^{2})\nonumber \\
 & \times\sum_{rs}\left\{ \theta(p^{0})\left[\bar{u}_{s}(\mathbf{p})\otimes u_{r}(\mathbf{p})\right]f_{rs}^{+}(x,\mathbf{p})\right.\nonumber \\
 & \left.-\theta(-p^{0})\left[\bar{v}_{s}(-\mathbf{p})\otimes v_{r}(-\mathbf{p})\right]f_{rs}^{-}(x,-\mathbf{p})\right\} \,,\label{eq:zeroth-order}
\end{align}
which can be obtained by using the quantized form of the free Dirac field
$\psi(x)$ and $\bar{\psi}(x)$ in Eq. (\ref{eq:definition_Wigner}).
The matrix-valued distributions $f_{rs}^{\pm}(x,\mathbf{p})$ are constructed by ensemble
averages of creation and annihilation operators for particles and
antiparticles \citep{Sheng:2019ujr,Sheng:2020oqs}.
The trace $\sum_s f_{ss}^{\pm}(x,\mathbf{p})$
is identified as the number density in phase space
for particles/antiparticles. A projection of $f_{rs}^{\pm}(x,\mathbf{p})$ onto Pauli matrices
gives the polarization density in the local rest frame. These distributions satisfy the kinetic equations
\begin{equation}
\bar{p}\cdot\partial f_{rs}^{\pm}(x,{\bf p})=0\,,
\label{eq:Boltzmann equation}
\end{equation}
where $\bar{p}^{\mu}\equiv(E_{{\bf p}},{\bf p})$ with $E_{{\bf p}}=\sqrt{{\bf p}^{2}+m^{2}}$.
Equations (\ref{eq:equation_Wigner}) and (\ref{eq:Boltzmann equation})
do not contain collision terms that have been derived in Ref. \citep{Yang:2020hri,Weickgenannt:2020aaf,Sheng:2021kfc,Wang:2021qnt}.
In this paper, collision terms are neglected because we focus on systems
in global or local thermal equilibrium.


The first and second order corrections in space-time gradient for
the Wigner function can be obtained by
solving the kinetic equation (\ref{eq:equation_Wigner}),
\begin{align}
\delta W(x,p) & =\frac{i}{4m}\left[\gamma^{\mu},\partial_{\mu}W_0(x,p)\right]\nonumber \\
 & +\frac{1}{16m^{2}}(\gamma\cdot\partial)W_0(x,p)(\gamma\cdot\overleftarrow{\partial})\nonumber \\
 & +\frac{\gamma\cdot p+m}{8m(p^{2}-m^{2})}\partial^{2}W_0(x,p)\,.
 \label{eq:first and second orders}
\end{align}
So the full solution of Eq. (\ref{eq:equation_Wigner}) is $W=W_0+\delta W$.
The expansion in space-time gradient is equivalent to that in the Planck constant $\hbar$.
The first order contribution agrees with the results in Refs. \citep{Gao:2019znl,Weickgenannt:2019dks,Hattori:2019ahi,Wang:2019moi}.

The appearance of $\delta W$ is a result of the uncertainty principle for quantum particles. If a system consists of point particles, the Wigner function will be given by $W_0$ only, with the momentum being constrained by the normal mass-shell condition  $p^2=m^2$. However, for quantum particles with finite space volume due to the constraint of the uncertainty principle, even though we neglect interactions among particles, the Wigner function contains corrections from non-local correlation, described by $\delta W$ in Eq. (\ref{eq:first and second orders}). These corrections include the electric dipole moment induced by an inhomogeneous charge distribution, the magnetization current, and the off-mass-shell correction. The first two corrections are widely known and can be found even in classical electrodynamics \citep{Jackson:1998nia}.
The off-mass-shell correction takes the form of a second order space-time derivative, which can also be obtained from the mass-shell equation of the Wigner function $(p^{2}-m^{2}-\partial^{2}/4)W(x,p)=0$. It has been recognized many years ago \citep{Elze:1986qd,Vasak:1987um} but is rarely studied when the system varies rapidly within the Compton wavelength, i.e., when $\text{Kn}\ll1$. In our power counting scheme, we identify them as second order corrections $\mathcal{O}(\text{Kn}^2)$.


In thermal equilibrium, we assume $f_{rs}^{\pm}(x,\mathbf{p})$ have
the following form as in Ref. \citep{Becattini:2013fla},
\begin{align}
f_{\text{eq},rs}^{+}(x,\mathbf{p})= & \frac{1}{2m}\bar{u}_{r}(\mathbf{p})\left(e^{\beta\cdot\bar{p}-\xi-\frac{1}{4}\omega^{\mu\nu}\sigma_{\mu\nu}}+1\right)^{-1}u_{s}(\mathbf{p})\,,\nonumber \\
f_{\text{eq},rs}^{-}(x,\mathbf{p})= & -\frac{1}{2m}\bar{v}_{r}(\mathbf{p})\left(e^{\beta\cdot\bar{p}+\xi+\frac{1}{4}\omega^{\mu\nu}\sigma_{\mu\nu}}+1\right)^{-1}v_{s}(\mathbf{p}),
\label{eq:equilibirum distributions}
\end{align}
where $\bar{p}^{\mu}\equiv(E_{{\bf p}},{\bf p})$ for both particles
and antiparticles, $\sigma^{\mu\nu}\equiv(i/2)[\gamma^\mu,\gamma^\nu]$ with $\gamma^\mu$ being gamma matrices. The parameters in Eq. (\ref{eq:equilibirum distributions}) are given by
$\beta^{\mu}\equiv\beta u^{\mu}$, $\xi\equiv\beta\mu$,
and $\omega^{\mu\nu}\equiv\beta\bar{\omega}^{\mu\nu}$,
with $\beta$, $u^{\mu}$, $\mu$, and $\bar{\omega}^{\mu\nu}$ being
the inverse temperature, the flow velocity, the chemical
potential, and the spin potential, respectively. In global equilibrium,
these thermodynamical parameters are space-time constants and then
Eq. (\ref{eq:equilibirum distributions}) can be derived rigorously
from quantum statistics \citep{Becattini:2013fla}.
We assume that in local equilibrium, the distributions have
the same expressions as those in global equilibrium but with space-time
dependent $\beta^{\mu}(x)$, $\xi(x)$, and $\omega^{\mu\nu}(x)$.

We note that the spin degrees of freedom are dissipative in the quark-gluon
plasma \citep{Li:2018srq,Li:2019qkf,Wang:2021qnt}. If the relaxation time of spin is much shorter than the typical
time scale of the fluid evolution, the considered system will be more
close to an instantaneous thermal equilibrium state in Eq. \ref{eq:equilibirum distributions}.
However the system can be off-equilibrium if the relaxation
time of spin is relatively large. The relaxation time of spin is recently
calculated in strongly coupled regime using the AdS/CFT correspondence \citep{Li:2018srq},
or the perturbative QCD \citep{Li:2019qkf}. In this study, we assume
that the spin relaxation is sufficiently fast, so the system is always in local equilibrium.
This is in analogy to the case of ideal hydrodynamics,
but equilibrium distributions in Eq. (\ref{eq:equilibirum distributions})
contain spin polarizations described by the spin potential $\bar{\omega}^{\mu\nu}$.

In later calculation, we focus on a weakly polarized system satisfying  $\chi_s\lesssim\text{Kn}\ll 1$. Then we can make expansions for thermodynamical parameters:
\begin{eqnarray}
\omega^{\mu\nu}&=&\omega^{(1)\mu\nu}+\omega^{(2)\mu\nu}+\cdots \,,\nonumber\\
\beta^{\mu}&=&\beta^{(0)\mu}+\beta^{(1)\mu}+\cdots\,,\nonumber\\
\xi&=&\xi^{(0)}+\xi^{(1)}+\cdots \,,
\end{eqnarray}
where `$(i)$' with $i=0,1,2,\cdots$ in superscripts denote
orders in $\text{Kn}$ and $\chi_s$. Note that the order of magnitude of $\chi_s$ is determined by $\omega^{\mu\nu}$ in thermal equilibrium, which will be shown in later calculations, therefore the spin potential is at least first order $\omega^{\mu\nu}\sim\mathcal{O}(\chi_s)$ and its space-time derivative $\partial^{\alpha}\omega^{\mu\nu}\sim\mathcal{O}(\text{Kn}\,\chi_s)$ is at least second order. We also expand the distributions in
Eq. (\ref{eq:equilibirum distributions}) up to second order in $\chi_s$ in Boltzmann limit,
\begin{align}
 & f_{\text{eq},rs}^{+}(x,\mathbf{p})=\exp(-\beta\cdot\bar{p}+\xi)\nonumber \\
 & \ \ \ \ \times\left[\left(1+\frac{1}{16}\omega^{\alpha\beta}\omega_{\alpha\beta}\right)
 \delta_{rs}+\frac{1}{8m}\omega^{\mu\nu}\bar{u}_{r}(\mathbf{p})\sigma_{\mu\nu}u_{s}(\mathbf{p})\right]\,,\nonumber \\
 & f_{\text{eq},rs}^{-}(x,\mathbf{p})=\exp(-\beta\cdot\bar{p}-\xi)\nonumber \\
 & \ \ \ \ \times\left[\left(1+\frac{1}{16}\omega^{\alpha\beta}\omega_{\alpha\beta}\right)
 \delta_{rs}+\frac{1}{8m}\omega^{\mu\nu}\bar{v}_{r}(\mathbf{p})\sigma_{\mu\nu}v_{s}(\mathbf{p})\right]\,.
 \label{eq:truncated distributions}
\end{align}
The Wigner function in local equilibrium can be obtained by
inserting Eq. (\ref{eq:truncated distributions}) into the zeroth
order Wigner function (\ref{eq:zeroth-order}) and higher order one
(\ref{eq:first and second orders}).


\vspace{1.5em}

\paragraph{Hydrodynamical quantities. \label{sec:Hydrodynamical-quantities}}

According to its definition, the Wigner function satisfies $W^{\dagger}=\gamma^{0}W\gamma^{0}$
and thus can be decomposed in terms of the generators of the Clifford
algebra $\Gamma_{i}=\{1,i\gamma^{5},\gamma^{\mu},\gamma^{5}\gamma^{\mu},\sigma^{\mu\nu}\}$,
\begin{equation}
W=\frac{1}{4}\left(\mathcal{F}+i\gamma^{5}\mathcal{P}+\gamma^{\mu}\mathcal{V}_{\mu}+\gamma^{5}\gamma^{\mu}\mathcal{A}_{\mu}+\frac{1}{2}\sigma^{\mu\nu}\mathcal{S}_{\mu\nu}\right)\,.
\label{Wigner_decomposition}
\end{equation}
where $1$ is the $4\times4$ unit matrix, $\gamma^{\mu}$ are gamma
matrices, $\gamma^{5}\equiv i\gamma^{0}\gamma^{1}\gamma^{2}\gamma^{3}$,
and $\sigma^{\mu\nu}\equiv (i/2)[\gamma^{\mu},\gamma^{\nu}]$.
The expansion coefficients $\mathcal{F}$, $\mathcal{P}$, $\mathcal{V}_{\mu}$,
$\mathcal{A}_{\mu}$, and $\mathcal{S}_{\mu\nu}$ are the scalar, pseudoscalar,
vector, axial vector, and tensor components of the Wigner function, respectively,
and they are all real \citep{Vasak:1987um}.
The current density $J^{\mu}(x)$, the energy-momentum tensor (density) $T^{\mu\nu}(x)$,
and the spin tensor (density) $S^{\lambda,\mu\nu}$ can be obtained from $\mathcal{V}_{\mu}$ and
$\mathcal{A}_{\mu}$ as follows
\begin{eqnarray}
J^{\mu}(x) &= & \int d^{4}p\,\mathcal{V}^{\mu}(x,p)\,, \label{curren-density}\\
T^{\mu\nu}(x) &= & \int d^{4}p\,p^{\nu}\mathcal{V}^{\mu}(x,p)\,,\label{eq:EMT}\\
S^{\lambda,\mu\nu}(x) &=& -\frac{1}{2}\epsilon^{\lambda\mu\nu\rho}\int d^{4}p\,\mathcal{A}_{\rho}(x,p)\,.
\label{eq:Spin tensor}
\end{eqnarray}
Note that when including spin degrees of freedom, the energy-momentum
tensor and the spin-tensor have the pseudo-gauge ambiguity \citep{Becattini:2012pp,Fukushima:2020qta,Speranza:2020ilk}.
Here we adopt the canonical definition in Eqs. (\ref{eq:EMT},\ref{eq:Spin tensor})
for $T^{\mu\nu}$ and $S^{\lambda,\mu\nu}$. Generally the canonical form of $T^{\mu\nu}$ contains
an antisymmetric part if the spin polarization is nonzero.


In local equilibrium, the Wigner function can be expressed in terms
of the distributions (\ref{eq:truncated distributions}).
It is straightforward to extract the vector component of the Wigner function
to obtain $J^{\mu}(x)$ and $T^{\mu\nu}(x)$ through Eqs. (\ref{curren-density},\ref{eq:EMT})
\begin{widetext}
\begin{align}
J_{\text{eq}}^{\mu}(x)= & \left(1+\frac{1}{16}\omega^{\alpha\beta}\omega_{\alpha\beta}\right)K_{1}u^{\mu}\sinh\xi-\frac{1}{16m^{2}}\partial^{2}\left[(2\beta K_{2}-5K_{1})u^{\mu}\sinh\xi\right]\nonumber \\
 & +\frac{1}{4m^{2}}\partial_{\nu}\left\{ \left[2 u^{[\mu}\omega_{\ \alpha}^{\nu]}u^{\alpha}(K_{2}+\beta^{-1}K_{1})+\omega^{\mu\nu}(K_{2}-\beta^{-1}K_{1})\right]\sinh\xi\right\} \,,\nonumber \\
T_{\text{eq}}^{\mu\nu}(x)= & \left(1+\frac{1}{16}\omega^{\alpha\beta}\omega_{\alpha\beta}\right)\left(u^{\mu}u^{\nu}K_{2}-\Delta^{\mu\nu}\beta^{-1}K_{1}\right)\cosh\xi+\frac{1}{4m^{2}}\partial_{\rho}\left(2\omega_{\alpha}^{\ [\mu}I^{\rho]\nu\alpha}+\omega^{\mu\rho}u^{\nu}m^{2}K_{1}\right)\cosh\xi\nonumber \\
 & +\frac{1}{16m^{2}}\partial^{2}\left\{ \left[g^{\mu\nu}\left(2K_{2}-5\beta^{-1}K_{1}\right)-u^{\mu}u^{\nu}\left(K_{2}+2\beta m^{2}K_{1}+\beta^{-1}K_{1}\right)\right]\cosh\xi\right\} \,,\label{eq:equilibrium J and Tmunu}
\end{align}
\end{widetext}
where $u^{\mu}$ is the flow velocity and $K_{n}$ are functions of $\beta$ as follows
\begin{equation}
K_{n}(\beta)\equiv\frac{8}{(2\pi)^{3}}\int\frac{d^{3}{\bf p}}{2E_{{\bf p}}}E_{{\bf p}}^{n}
e^{-\beta E_{{\bf p}}}\,.\label{eq:functions Kn}
\end{equation}
They satisfy the following recursive relation
\begin{equation}
K_{n}=\frac{n+1}{\beta}K_{n-1}+m^{2}K_{n-2}-\frac{n-2}{\beta}m^{2}K_{n-3}\,,\label{eq:recurrence-1}
\end{equation}
which allows us to express $K_{n}$ in terms of any $K_{i}$
and $K_{j}$ with $i\neq j$. In Eq. (\ref{eq:equilibrium J and Tmunu})
and in the following, we choose $K_{1}$ and $K_{2}$ as basis functions
to express $K_{n}$, since $K_{1}$ and $K_{2}$ are related
to the particle number and energy density when neglecting
second order terms, while other $K_{n}$ do not
have explicit physical meanings.
The rank-3 moment $I^{\mu\nu\alpha}$ in $T_{\text{eq}}^{\mu\nu}$
in Eq. (\ref{eq:equilibrium J and Tmunu}) is given by
\begin{eqnarray}
I^{\mu\nu\alpha}&=&u^{\mu}u^{\nu}u^{\alpha}K_{3}+\frac{1}{3}(\Delta^{\mu\nu}u^{\alpha}\nonumber\\
&&+\Delta^{\mu\alpha}u^{\nu}+\Delta^{\nu\alpha}u^{\mu})(m^{2}K_{1}-K_{3})\,,
\end{eqnarray}
where the projection operator is defined as $\Delta^{\mu\nu}\equiv g^{\mu\nu}-u^{\mu}u^{\nu}$.


One can see in Eq. (\ref{eq:equilibrium J and Tmunu}) that the spin dependent terms
are at least second order in the space-time gradient.
At zeroth order in $\text{Kn}$ and $\chi_s$,
$J_{\text{eq}}^{\mu}$ and $T_{\text{eq}}^{\mu\nu}$
agree with the results in ideal hydrodynamics \citep{Romatschke:2017ejr}. With the help of the fluid velocity $u^\mu$,
equation (\ref{eq:equilibrium J and Tmunu}) can be rewritten as
\begin{eqnarray}
J_{\text{eq}}^{\mu} &= & n_{\text{eq}}u^{\mu}+\delta j^{\mu}\,,\nonumber \\
T_{\text{eq}}^{\mu\nu} &= & \epsilon_{\text{eq}}u^{\mu}u^{\nu}-P_{\text{eq}}\Delta^{\mu\nu}+\delta T_{S}^{\mu\nu}+\delta T_{A}^{\mu\nu}\,, \label{constitutional-relation}
\end{eqnarray}
where $n_{\text{eq}}$, $\epsilon_{\text{eq}}$, and $P_{\text{eq}}$
are the particle number density, the energy density, and the pressure respectively,
and they can be extracted as
\begin{widetext}
\begin{eqnarray}
n_{\text{eq}} & \equiv & u_{\mu}J_{\text{eq}}^{\mu} = \left(1+\frac{1}{16}\omega^{\alpha\beta}\omega_{\alpha\beta}\right)
K_{1}\sinh\xi+\frac{1}{4m^{2}}(\nabla_{\mu}u_{\nu})\omega^{\mu\nu}(K_{2}+\beta^{-1}K_{1})\sinh\xi
\nonumber \\
&&-\frac{1}{2m^{2}}u_{\mu}\nabla_{\nu}\left[\omega^{\mu\nu}\beta^{-1}K_{1}\sinh\xi\right]-\frac{1}{16m^{2}}\partial^{2}
\left[(2\beta K_{2}-5 K_{1})\sinh\xi\right]\,,
\nonumber \\
\epsilon_{\text{eq}}&\equiv & u_{\mu}u_{\nu}T_{\text{eq}}^{\mu\nu} = \left(1+\frac{1}{16}\omega^{\alpha\beta}\omega_{\alpha\beta}\right)
K_{2}\cosh\xi+\frac{1}{4m^{2}}(\nabla_{\mu}u_{\nu})\omega^{\mu\nu}
\left[5\beta^{-1}(K_{2}+\beta^{-1}K_{1})+m^{2}K_{1}\right]\cosh\xi
\nonumber \\
&&-\frac{1}{2m^{2}}u_{\mu}\nabla_{\nu}\left[\omega^{\mu\nu}\beta^{-1}(K_{2}+\beta^{-1}K_{1})\cosh\xi\right]
+\frac{1}{16m^{2}}\partial^{2}\left[\left(K_{2}-6\beta^{-1}K_{1}-2\beta m^{2}K_{1}\right)\cosh\xi\right]\,,\label{eq:equilibrium n and e}
\end{eqnarray}
\end{widetext}
where $\nabla^{\mu}\equiv\Delta^{\mu\nu}\partial_{\nu}$, and the pressure is
obtained by $P_{\text{eq}}\equiv -(1/3) \Delta_{\mu\nu}T_{\text{eq}}^{\mu\nu}$,
giving the trace anomaly
\begin{eqnarray}
\epsilon_{\text{eq}}-3P_{\text{eq}} & = & \left(1+\frac{1}{16}\omega^{\alpha\beta}\omega_{\alpha\beta}\right)m^{2}K_{0}\cosh\xi\nonumber \\
&& +\frac{1}{16}\partial^{2}\left[\left(7K_{0}-2\beta K_{1}\right)\cosh\xi\right]\,.
 \label{eq:trace anomaly}
\end{eqnarray}
The residue terms $\delta j^{\mu}$, $\delta T_{S}^{\mu\nu}$, and
$\delta T_{A}^{\mu\nu}$ in Eq. (\ref{constitutional-relation})
describe deviations from ideal hydrodynamics without spin.
They are at least second order in the space-time gradient.
We identify $\delta j^{\mu}$ as the magnetization current induced
by the inhomogeneity of the spin polarization. Here $\delta T_{S}^{\mu\nu}$
and $\delta T_{A}^{\mu\nu}$ are symmetric and antisymmetric respectively for
an interchange of $\mu$ and $\nu$.
It is straightforward to read out
the explicit expressions for $\delta j^{\mu}$, $\delta T_{S}^{\mu\nu}$, and $\delta T_{A}^{\mu\nu}$
from Eq. (\ref{eq:equilibrium J and Tmunu}).
They contain the heat flow and viscous tensor correction
induced by the spin polarization that have been discussed in the
quantum spin vorticity theory \citep{RN100}.
In our power counting scheme, these spin-induced terms are of order $\mathcal{O}(\text{Kn}\,\chi_s)$, which is much smaller than first-order viscous corrections, and  comparable to second-order corrections in regular viscous hydrodynamics without spin. That is because
in the quark-gluon plasma produced in heavy-ion collisions, the average spin polarization of particles would be $\sim0.02$, as indicated by the $\Lambda$'s global polarization \citep{STAR:2017ckg,Adam:2018ivw}. The weakly polarized condition $\chi_s\lesssim\text{Kn}$ therefore can be fulfilled ($\text{Kn}$ in heavy-ion collisions has been calculated in Ref. \citep{Niemi:2014wta}). Note that $\delta j^{\mu}$ and $\delta T_{S}^{\mu\nu}$ also contain second order terms $O(\text{Kn}^2)$. These terms are contributions from long-range correlations, which do not appear in regular ideal hydrodynamics.


Even through we deal with hydrodynamics in local equilibrium,
we still obtain viscous terms in Eq. (\ref{eq:equilibrium J and Tmunu}).
Similar to the viscous hydrodynamics, the definition of the rest frame
and $u^{\mu}$ is subtle because the charge and energy currents
are in general not parallel when we include higher order corrections.
One can either choose the Eckart frame with $u_{E}^{\mu}\equiv J_{\text{eq}}^{\mu}/\sqrt{J_{\text{eq}}^{\nu}J_{\text{eq},\nu}}$,
or the Landau frame with $T_{\text{eq}}^{\mu\nu}u_{L,\nu}=\epsilon_{\text{eq}}u_{L}^{\mu}$.
The difference between these two choices is in the second order in $\text{Kn}$ and $\chi_s$, i.e., $u_{E}^{\mu}-u_{L}^{\mu}\simeq\mathcal{O}(\text{Kn}^2,\,\text{Kn}\,\chi_s,\,\chi_s^2)$.


For the angular momentum, we look at the spin tensor $S^{\lambda,\mu\nu}(x)$
and the dipole-moment tensor $M^{\mu\nu}(x)$. The spin tensor $S^{\lambda,\mu\nu}(x)$
is one part of the total angular momentum tensor $J^{\lambda,\mu\nu}(x)=x^{\mu}T^{\lambda\nu}-x^{\nu}T^{\lambda\mu}+S^{\lambda,\mu\nu}$.
Note that the separation of spin and orbital angular momentum is not unique
and subject to pseudogauge transformation \citep{Becattini:2012pp,Fukushima:2020qta,Speranza:2020ilk}.
Here we adopt the canonical definition for $S^{\lambda,\mu\nu}(x)$ as in Eq. (\ref{eq:Spin tensor}).
In local equilibrium, we can express it in terms of $\beta$, $\xi$, $u^{\mu}$,
and $\omega^{\mu\nu}$ as
\begin{equation}
S_{\text{eq}}^{\lambda,\mu\nu}(x)=\frac{1}{4}\left(u^{\lambda}\omega^{\mu\nu}+2u^{[\mu}\omega^{\nu]\lambda}\right)K_{1}\cosh\xi\,,\label{eq:equilibrium S}
\end{equation}
where $K_{1}$ is defined in Eq. (\ref{eq:functions Kn}).
The spin potential $\omega^{\mu\nu}$ contains six independent variables.
However, $S_{\text{eq}}^{\lambda,\mu\nu}$ only contains three independent variables
because it only depends on $\tilde{\omega}^{\mu}\equiv\epsilon^{\mu\nu\alpha\beta}u_{\nu}\omega_{\alpha\beta}/2$,
\begin{equation}
S_{\text{eq}}^{\lambda,\mu\nu}(x)=\frac{1}{4}\epsilon^{\lambda\mu\nu\beta}\tilde{\omega}_{\beta}K_{1}\cosh\xi\,.
\end{equation}
In order to study the remaining three degrees of freedom of $\omega^{\mu\nu}$,
we look at the dipole momentum tensor defined as
\begin{equation}
D^{\mu\nu}(x)=\int d^{4}p\,\mathcal{S}^{\mu\nu}(x,p)\,,
\end{equation}
where $\mathcal{S}^{\mu\nu}$ is the tensor component of the Wigner function (\ref{Wigner_decomposition}). The dipole moment tensor $D^{\mu\nu}$ is antisymmetric with respect to its indices,
which allows the following decomposition
\begin{equation}
D^{\mu\nu}=\mathcal{E}^{\mu}u^{\nu}-\mathcal{E}^{\nu}u^{\mu}
-\epsilon^{\mu\nu\alpha\beta}u_{\alpha}\mathcal{M}_{\beta}\,.
\end{equation}
Such a decomposition depends on the choice of the frame velocity $u^{\mu}$.
We identify $\mathcal{E}^{\mu}=D^{\mu\nu}u_{\nu}$ as the electric dipole
moment vector and $\mathcal{M}^{\mu}=-(1/2) \epsilon^{\mu\nu\alpha\beta}u_{\nu}D_{\alpha\beta}$
as the magnetic dipole moment vector, because they are coupled with electric field and magnetic field respectively if we consider an external electromagnetic field as in Ref. \cite{Sheng:2020oqs}.
In local equilibrium, we obtain
\begin{align}
\mathcal{E}_{\text{eq}}^{\mu}= & -\frac{1}{m\beta}\left(\omega^{\mu\nu}u_{\nu}+\frac{1}{2}\beta u_{\nu}\partial^{\nu}u^{\mu}\right)K_{1}\sinh\xi\nonumber \\
 & +\frac{1}{2m}\nabla^{\mu}\left(K_{1}\sinh\xi\right)\,,\nonumber \\
\mathcal{M}_{\text{eq}}^{\mu}= & -\frac{1}{2m}\tilde{\omega}^{\mu}(K_{2}-\beta^{-1}K_{1})\sinh\xi\nonumber \\
 & -\frac{1}{2m}\epsilon^{\mu\nu\alpha\beta}u_{\nu}(\partial_{\alpha}u_{\beta})K_{1}\sinh\xi\,.
\end{align}
The magnetic dipole moment consists of two parts, one is proportional
to $\tilde{\omega}^{\mu}$, which is the contribution of intrinsic spin degrees of freedom;
the other part arises from the rotation of the fluid or the orbital angular momentum
proportional to the vorticity $\epsilon^{\mu\nu\alpha\beta}u_{\nu}(\partial_{\alpha}u_{\beta})$.
The electric dipole moment, on the other hand, has three parts: the
electric dipole moment of particles, the acceleration of the fluid,
and the contribution of the inhomogeneous particle distribution, which are proportional
to $\omega^{\mu\nu}u_{\nu}$, $u_{\nu}\partial^{\nu}u^{\mu}$, and
$\nabla^{\mu}n_{\text{eq}}$, respectively.
In local equilibrium, a non-vanishing electric dipole moment is allowed.


\vspace{1.5em}

\paragraph{EOMs for thermodynamical parameters. \label{sec:EOMs}}

For the ideal hydrodynamics without spin, the current density $J^{\mu}(x)$
and the energy-momentum tensor $T^{\mu\nu}(x)$ are conserved respectively,
\begin{align}
\partial_{\mu}J^{\mu}(x)= & 0\,,\nonumber \\
\partial_{\nu}T^{\mu\nu}(x)= & 0\,.\label{eq:conservation laws}
\end{align}
These conservation laws still hold in the presence of spin degrees
of freedom. Inserting the equilibrium expressions of $J^{\mu}(x)$
and $T^{\mu\nu}(x)$ in Eq. (\ref{eq:equilibrium J and Tmunu}) into
the conservation laws (\ref{eq:conservation laws}), we derive the
equations of motions for $\beta$, $\xi$, and $u^{\mu}$,
\begin{align}
\dot{\beta}= & \frac{K_{2}+\beta^{-1}K_{1}\cosh^{2}\xi}{K_{1}K_{3}\cosh^{2}\xi-K_{2}K_{2}\sinh^{2}\xi}K_{1}\theta\,,\nonumber \\
\dot{\xi}= & \frac{\left(K_{2}+\beta^{-1}K_{1}\right)K_{2}-K_{1}K_{3}}{K_{1}K_{3}\cosh^{2}\xi-K_{2}K_{2}\sinh^{2}\xi}\theta\sinh\xi\cosh\xi\,,\nonumber \\
\dot{u}^{\mu}= & \frac{K_{1}}{K_{1}+\beta K_{2}}\tanh\xi\nabla^{\mu}\xi-\frac{1}{\beta}\nabla^{\mu}\beta\,,\label{eq:EOMs 1}
\end{align}
where dot represents the time derivative in comoving frame defined
as $d/d\tau\equiv u_{\mu}\partial^{\mu}$, and $\nabla^{\mu}\equiv\Delta^{\mu\nu}\partial_{\nu}$
is the space derivative. With the help of Eqs. (\ref{eq:equilibrium n and e})
and (\ref{eq:trace anomaly}), we find that
$K_{1}\simeq n_{\text{eq}}/\sinh\xi\simeq\beta P_{\text{eq}}/\cosh\xi$,
$K_{2}\simeq\epsilon_{\text{eq}}/\cosh\xi$, with corrections to $K_1$ and $K_2$ are at least second order in $\text{Kn}$ and $\chi_s$,
and thus $K_{3}=\frac{3}{\beta}\left(K_{2}+\beta^{-1}K_{1}\right)+m^{2}K_{1}$
according to the recurrence relation (\ref{eq:recurrence-1}). Therefore,
equations in (\ref{eq:EOMs 1}) can be expressed in term of the equilibrium
quantities $n_{\text{eq}}$, $P_{\text{eq}}$, and $\epsilon_{\text{eq}}$.
Once we do this, we will find that a nonzero $\omega^{\mu\nu}$ does
not contribute at the leading and next-to-leading order in $\text{Kn}$ and $\chi_s$. Such a conclusion
can be obtained from another point of view from $J_{\text{eq}}^{\mu}$
and $T_{\text{eq}}^{\mu\nu}$: the deviation from ideal hydrodynamics
without spin is at least $\mathcal{O}(\text{Kn}^2,\,\text{Kn}\,\chi_s,\,\chi_s^2)$.

On the other hand, the conservation of the total angular momentum gives
\begin{equation}
\partial_{\lambda}S^{\lambda,\mu\nu}(x)=T^{\nu\mu}(x)-T^{\mu\nu}(x)\,.
\end{equation}
Inserting equilibrium expressions in Eqs. (\ref{eq:equilibrium J and Tmunu},\ref{eq:equilibrium S})
into the above equation, we obtain the
equation of motion for $\omega^{\mu\nu}$ in $\dot{\omega}^{\mu\nu}$, which can be
decomposed into
\begin{equation}
\dot{\omega}^{\mu\nu}=\Delta_{\alpha}^{\mu}\Delta_{\beta}^{\nu}\dot{\omega}^{\alpha\beta}
-u^{\mu}\dot{\omega}^{\nu\alpha}u_{\alpha}+u^{\nu}\dot{\omega}^{\mu\alpha}u_{\alpha}\,,
\label{eq:EOM omega}
\end{equation}
where the first term and $\dot{\omega}^{\mu\nu}u_{\nu}$ appearing in last two terms are given by
\begin{eqnarray}
\Delta_{\alpha}^{\mu}\Delta_{\beta}^{\nu}\dot{\omega}^{\alpha\beta} & = & C_{3}\Delta_{\alpha}^{\mu}\Delta_{\beta}^{\nu}\omega^{\alpha\beta}+C_{4}\Delta_{\beta}^{[\mu}\sigma_h^{\nu]\rho}
\omega_{\ \rho}^{\beta}\nonumber \\
&& -\frac{1}{2}C_{4}(\nabla^{[\mu}\omega_{\ \rho}^{\nu]})u^{\rho}
+C_{2}C_{4}u^{\rho}\omega_{\ \rho}^{[\mu}\nabla^{\nu]}\xi \,,\nonumber \\
\dot{\omega}^{\mu\nu}u_{\nu} & = & C_{1}\omega^{\mu\nu}u_{\nu}
+C_{2}\Delta_{\rho}^{\mu}\omega^{\rho\nu}\nabla_{\nu}\xi \nonumber \\
&& +\sigma_h^{\mu\nu}\omega_{\nu\rho}u^{\rho}+\frac{1}{2}\Delta_{\rho}^{\mu}(\nabla^{\nu}\omega_{\ \nu}^{\rho})\,,
\end{eqnarray}
where $\sigma_h^{\mu\nu}\equiv(\nabla^\mu u^\nu+\nabla^\nu u^\mu)/2-\Delta^{\mu\nu}\theta/3$. The transport coefficients $C_{i}$ ($i=1,2,3,4$) are given by
\begin{align}
C_{1}= & -\frac{5}{3}\theta+\frac{5}{\beta}\dot{\beta}+\frac{\beta m^{2}K_{1}}{K_{1}+\beta K_{2}}\dot{\beta}-\tanh\xi\ \dot{\xi}\,,\nonumber \\
C_{2}= & \frac{1}{2}\left(1-\frac{5K_{1}}{K_{1}+\beta K_{2}}-\frac{m^{2}\beta^{2}K_{1}K_{1}}{(K_{1}+\beta K_{2})^{2}}\right)\tanh\xi\,,\nonumber \\
C_{3}= & -\left[5-\frac{2\beta^{2}m^{2}K_{1}}{2\left(K_{1}+\beta K_{2}\right)+\beta^{2}m^{2}K_{1}}\right]\frac{\theta}{3}\nonumber \\
 & -\tanh\xi\ \dot{\xi}+\left[5-\frac{m^{2}\beta^{2}(3K_{1}-\beta K_{2})}{2(K_{1}+\beta K_{2})+\beta^{2}m^{2}K_{1}}\right]\frac{\dot{\beta}}{\beta}\,,\nonumber \\
C_{4}= & \frac{\beta^{2}m^{2}K_{1}}{2(K_{1}+\beta K_{2})+\beta^{2}m^{2}K_{1}}-1\,.
\end{align}
In our power counting scheme, we have $\omega^{\mu\nu}\sim\mathcal{O}(\chi_s)$
and thus both sides of Eq. (\ref{eq:EOM omega}) are
$\mathcal{O}(\text{Kn}\,\chi_s)$.
Equation (\ref{eq:EOM omega}) has a trivial solution $\omega^{\mu\nu}=0$,
corresponding to the spinless case. Note that similar results have
been derived in Ref. \citep{Bhadury:2020cop}, but the transport coefficients
are slightly different. That is because we include spin-induced energy
current, which leads to an antisymmetric part in $T_{\text{eq}}^{\mu\nu}$,
while $T_{\text{eq}}^{\mu\nu}$ used in Ref. \citep{Bhadury:2020cop} is symmetric.

\vspace{1.5em}

\paragraph{Summary. \label{sec:Summary}}

In this paper, we formulate ideal hydrodynamics with spin degrees
of freedom from the Wigner function approach.
A general solution to the kinetic equation for the Wigner function has been found
up to the second order in space-time gradient.
The solution contains a zeroth order part $W_0$ and a high-order correction part $\delta W$.
The zeroth order part $W_0$
can be expressed in terms of matrix-valued
distributions, while $\delta W$ can be expressed by space-time
derivatives of $W_0$. We adopt the equilibrium form of the matrix-valued distributions
for particles and antiparticles proposed in Ref. \citep{Becattini:2013fla}.

Based on the Wigner function in local equilibrium, we derive
hydrodynamical quantities: the current density, the energy-momentum
tensor, the spin tensor, and the electric/magnetic dipole moment.
Comparing with ideal hydrodynamics without spin, the current density
and the energy-momentum tensor contain additional contributions at the second order in the Knudsen number $\text{Kn}$ and the average spin polarization $\chi_s$ from space-time gradients and from spin degrees of freedom respectively.
We also find that the spin tensor and magnetic dipole moment depend on
$\epsilon^{\mu\nu\alpha\beta}u_{\nu}\omega_{\alpha\beta}$
but does not depend on $\omega^{\mu\nu}u_{\nu}$, while the electric
dipole moment is related to $\omega^{\mu\nu}u_{\nu}$. We also derive
equations of motions for thermodynamical parameters $\beta$,
$u^{\mu}$, $\xi$, and $\omega^{\mu\nu}$ from conservation laws.
These equations can be applied to numerical simulations for the spin
evolution in the quark-gluon plasma. We emphasize that this work is based on
the assumption of local equilibrium.
In more realistic cases, whether the relaxation time of spin is short
enough or not is still under debate.
On the other hand, ideal hydrodynamics without spin is acausal and not well-defined, and ideal spin hydrodynamics may also have similar problems \citep{Speranza:2021bxf}. To avoid these problems, dissipative effects induced by particle collisions have to be included.

\vspace{1.5em}

\begin{acknowledgments}
The authors thank Long-Gang Pang and Shi Pu for helpful discussions.
H.-H. P. and Q.W. are supported in part by the National Natural Science
Foundation of China (NSFC) under Grants 11890713 (a sub-grant of 11890710)
and 11947301, and by the Strategic Priority Research Program of Chinese
Academy of Sciences under Grant XDB34030102. X.-L. S. is supported
by the National Natural Science Foundation of China (NSFC) under grants
11935007, 11221504, 11861131009, 11890714 (a sub-grant of 11890710)
and 12047528.
\end{acknowledgments}

\bibliographystyle{apsrev}
\bibliography{ideal_spin_hydro}

\end{document}